\documentclass[
  aps,
  prb,
  twocolumn,
  superscriptaddress]{revtex4-2}
\usepackage{amsmath,amssymb,bm}
\usepackage{graphicx}
\usepackage{dcolumn}
\usepackage{color}
\usepackage{accents}
\usepackage{mathtools}
\usepackage{ulem}
\usepackage{url}
\usepackage{hyperref}

\begin{document}

\title{Thermoelectric effect in a superconductor with Bogoliubov Fermi surfaces}
\date{\today}

\author{Tomoya Sano}
\affiliation{Department of Applied Physics, Hokkaido University, Sapporo 060-8628, Japan}
\author{Takumi Sato}
\affiliation{Department of Applied Physics, Hokkaido University, Sapporo 060-8628, Japan}
\affiliation{Graduate School of Science, Hokkaido University, Sapporo 060-0810, Japan}
\author{Akihiro Sasaki}
\affiliation{Department of Applied Physics, Hokkaido University, Sapporo 060-8628, Japan}
\author{Satoshi Ikegaya}
\affiliation{Department of Applied Physics, Nagoya University, Nagoya 464-8603, Japan}
\affiliation{Institute for Advanced Research, Nagoya University, Nagoya 464-8601, Japan}
\author{Shingo Kobayashi}
\affiliation{RIKEN Center for Emergent Matter Science, Wako, Saitama 351-0198, Japan}
\author{Yasuhiro Asano}
\affiliation{Department of Applied Physics, Hokkaido University, Sapporo 060-8628, Japan}

\begin{abstract}
We study theoretically the thermoelectric effect in a superconducting state having 
the Bogoliubov-Fermi surfaces which stays in a thin superconducting layer 
between a conventional superconductor and an insulator.
The thermoelectric coefficients calculated based on the linear response theory 
show the remarkable anisotropy in real space, which are explained well 
by the anisotropic shape of the Bogoliubov-Fermi surface in momentum space.
Our results indicate a way to check the existence of the Bogoliubov-Fermi surfaces 
in a stable superconducting state because the anisotropy is controlled by the 
direction of an applied magnetic field.
\end{abstract}

\maketitle

\section{Introduction}
In a superconductor belonging to unconventional symmetry class such as 
$d$-wave and $p$-wave, a Bogoliubov quasiparticle 
at zero energy (Fermi level) exists only at nodes of the pair potential.
This conclusion is valid when the superconducting states are described 
effectively by the $2 \times 2$ Bogoliubov-de Gennes (BdG) Hamiltonian.
When an electron has internal degrees of freedom, such as spin and sublattice, 
the size of the BdG Hamiltonian becomes as large as $4 \times 4$ or more.
A Bogoliubov quasiparticle in such superconducting states can form Fermi surfaces called 
Bogoliubov-Fermi surfaces (BFSs)~\cite{volovik1989zeroes,Yang:prb1998,Agterberg2017BFS}.
Therefore, multiband superconductors, $j=3/2$ superconductors, and usual $s=1/2$ superconductors
with spin-dependent potentials can have the BFSs in their superconducting states~\cite{Burset:prb2015,Agterberg2017BFS,Timm:prb2017,Bzdusek:prb2017,YuanFu2018Zeemaninduced,Brydon2018GeneralBFS,
Menke2019stanilizedbySOC,Julia:prl2020,setty2020BFSspin1/2,Timm:prb2021,Kobayashi:prb2022,Ohashi2024tunneling,Serafim:Scipost2024}.

The residual finite density of states (DOS) at zero energy is a direct consequence of the BFSs, 
which can be observed by using spectroscopies on a surface~\cite{Hanaguri:Sci2018,Nagashima:RS2022} 
and modifies the bulk properties such as
specific heat~\cite{setty2020BFSspin1/2,setty2020topological,Timm2020ExpofBFS},
magnetic properties, and transport 
properties~\cite{Sato:pnas2018,Mizukami2023communphys,Zhongyu:CommPhys2023,Nagashima:prl2024}.
However, it is difficult to catch convincing evidences of the BFSs because the 
residual density of states is derived also from random impurities.
Theoretical studies have suggested to catch signatures of the BFSs
through transport properties in a normal-metal/superconductor 
junction ~\cite{setty2020BFSspin1/2,setty2020topological,Timm2020ExpofBFS,Zhu2021Discovery,Banerjee2022Fano,Mateos:prb2024,Ohashi2024tunneling,Pal:prb2024}.
It should be noted that superconducting states having BFSs are thermodynamically 
unstable as pointed out by numerical simulations for $j=3/2$ 
superconductors~\cite{Menke2019stanilizedbySOC,Bhattacharya:prb2023}.
The instability is partially derived from a fact that 
quasiparticles on the BFSs coexist with odd-frequency 
Cooper pairs~\cite{Kim2021BFSandodd-w,asano:prb2015,sato:prb2024}.
A possibility of an odd-frequency superconductivity has been discussed as a 
result of Cooper pairing of two quasiparticles on the BFSs~\cite{miki:prb2021}. 
In order to establish the physics of quasiparticles on the BFSs, it is necessary to 
realize stable superconducting states with BFSs and clarify their specific phenomena.
In what follows, we address these issues.

In this paper, we theoretically discuss the thermoelectric effect due to the quasiparticle on the BFSs.
The electric current $\bm{j}$ flows in the presence of the spatial 
gradient of a temperature $\nabla T$.  
In the relation $\bm{j} = - \alpha \nabla T$,
the thermoelectric coefficient $\alpha$ represets the strength of the effect.
The thermoelectric coefficient in a uniform of superconductor $\alpha_{\mathrm{S}}$
has been formulated in terms of the solutions of
 the transport equation~\cite{galperin1973nonlinear,galperin1974thermoelectric}
and the quasiclassical Green's functions~\cite{Graf1996Electtic_thermal}.
The expression at low temperatures $\alpha_{\mathrm{S}} \approx  \alpha_{\mathrm{N}}
\, \exp(- \Delta/T) $ suggests that the thermoelectric coefficient of a conventional superconductor  
is exponentially smaller than that in the normal state $\alpha_{\mathrm{N}}$.
The results can be explained well by the absence of the DOS around zero energy
due to the pair potential $\Delta$.
We discuss the effects of a quasiparticle on the BFSs on the thermoelectric 
effect in a stable superconducting state at a semiconductor thin film 
as illustrated in Fig.~\ref{fig:system}.
The BFSs appear at the film in the presence of both spin-orbit interactions (SOI) 
and strong magnetic fields~\cite{Burset:prb2015,YuanFu2018Zeemaninduced,Serafim:Scipost2024}.
The thermoelectric coefficient is calculated based on the linear response theory
by using the Keldysh Green's function method.
The calculated results show that the thermoelectric effect is 
anisotropic in real space, which reflects the anisotropic shape of the BFSs in momentum space. 
Moreover, the thermoelectric coefficient in the presence of the BFSs can be larger than 
its normal state value. The enhancement in the thermoelectric effect is explained by 
the shift of the gapped DOS due to a magnetic field.
We conclude that the existence of the BFSs can be directly confirmed by the anisotropy of the 
thermoelectric coefficient.

This paper is organized as follows. 
In Sec.~\ref{Sec2}, we describe the realization of BFSs in the semiconductor-superconductor hybrid system.
In Sec.~\ref{sec:formula}, we derive the thermoelectric coefficient within the linear response to the spatial gradient of temperature.
In Sec.~\ref{Sec3}, the calculated results of the DOS 
and those of the thermoelectric coefficients are presented. 
We discuss the meaning of the obtained results in Sec.~\ref{Sec4}.
The conclusion is given in Sec.~\ref{Sec5}.

\section{Model}\label{Sec2}
\begin{figure}[ttt]
  \includegraphics[width=7cm]{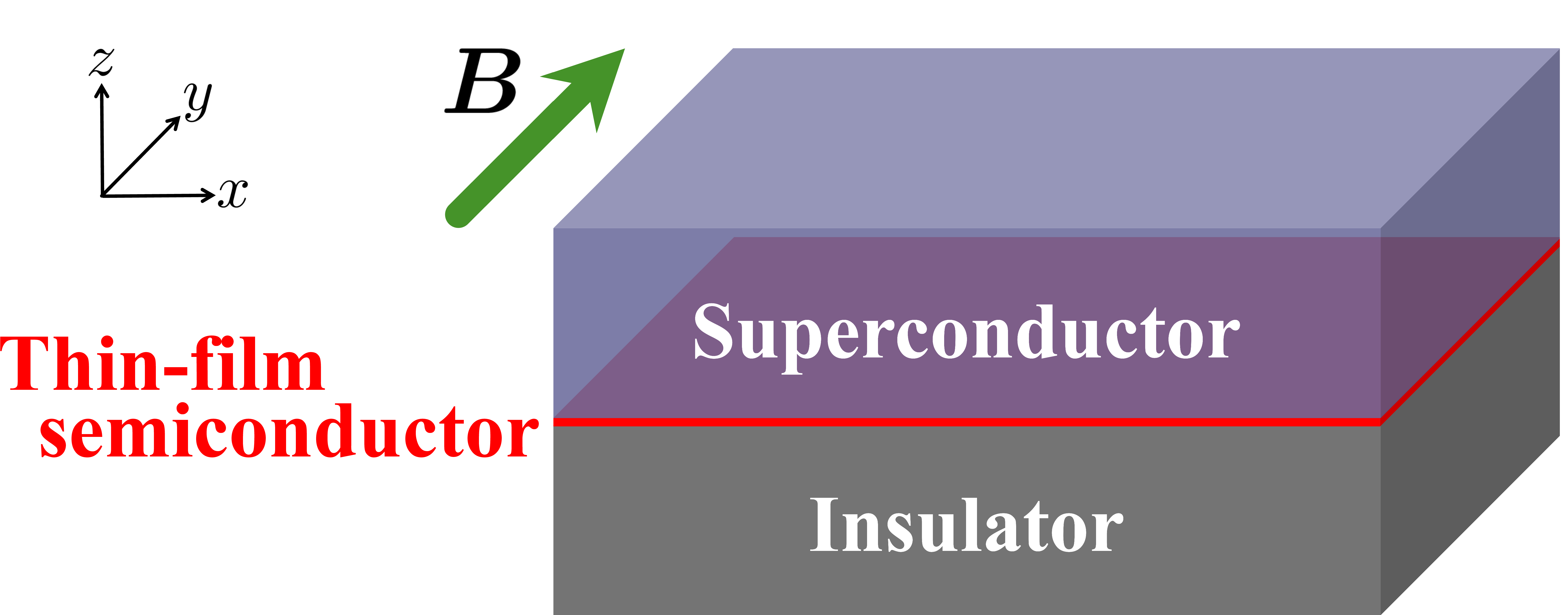}
  \caption{Schematics of a proximity structure considered in this paper. 
  We focus on two-dimensional electronic states at a thin semiconductor sandwiched
  between a superconductor and an insulator.
  Because of inversion symmetry breaking in the $z$ direction, Rashba spin-orbit interaction
  works for electrons.
  In the presence of an external magnetic field applied in the $y$ direction and the
  proximity-effect induced pair potential, a superconducting 
  state with Bogoliubov-Fermi surfaces can be realized in the thin film.
  }
  \label{fig:system}
\end{figure}

We discuss the thermoelectric effect in two-dimensional electronic 
states realized in a thin semiconductor sandwiched 
by a spin-singlet $s$-wave superconductor and an insulator 
under an in-plane magnetic field as illustrated in 
Fig.~\ref{fig:system}. 
The BdG Hamiltonian in a thin semiconductor~\cite{Serafim:Scipost2024,YuanFu2018Zeemaninduced} reads
\begin{align}
  \check{H}_{\mathrm{BdG}}
  &=
  \begin{bmatrix}
    \hat{h}_{\mathrm{N}}(\bm{k}) & \Delta i \hat{\sigma}_y \\
    - \Delta i \hat{\sigma}_y & - \hat{h}_{\mathrm{N}}^\ast(-\bm{k})
  \end{bmatrix} , \label{eq:BdG} \\
  \hat{h}_{\mathrm{N}} (\bm{k})
  &=
  \xi_{\bm{k}} \hat{\sigma}_0 - \bm{V} \cdot \hat{\bm{\sigma}}
  - \bm{\lambda} \times \bm{k} \cdot \hat{\bm{\sigma}}, \label{eq:hn}\\
  \xi_{\bm{k}} =& \hbar^2 \bm{k}^2/(2m) - \mu, \quad 
  \bm{V} = (0, V, 0),
\end{align}
with $V=\frac{1}{2} g \mu_B B$, 
where $\mu$ is the chemical potential, $B$ is the amplitude of an external magnetic 
field in the $y$ direction, 
$g$ is the $g$-factor, and $\mu_B$ is the Bohr magneton. 
The Pauli's matrices in spin space are denoted by  $\hat{\bm{\sigma}} = (\hat{\sigma}_x, 
\hat{\sigma}_y, \hat{\sigma}_z)$ 
and $\hat{\sigma}_0$ is the unit matrix in spin space. 
The pair potential at the thin film is $\Delta$.
The vector $\bm{\lambda}$ is parallel to the $z$ direction
because inversion symmetry is broken along the $z$ direction in
 the proximity structure. 
In such a situation, a Zeeman field in the $xy$ plane enables the Bogoliubov-Fermi surfaces~\cite{YuanFu2018Zeemaninduced,Banerjee2022Fano,Serafim:Scipost2024,Burset:prb2015}.
The reasons why we need to consider such a junction are explained below.

It is widely accepted that spin-singlet superconductivity is fragile in strong Zeeman
potential $V$. 
However, to realize the BFSs in a conventional 
superconductor~\cite{YuanFu2018Zeemaninduced,Burset:prb2015}, 
the amplitude of a Zeeman field must be larger than 
the Clogston-Chandrasekhar limit~\cite{chandrasekhar:apl1962,clogston:prl1962}.
Here, we assume that
the pair potential $\Delta$ is induced in a thin semiconductor due to the proximity effect
from a parent superconductor.
In such a system, induced pair potential would be smaller than that of the parent superconductor $\Delta_{\mathrm{bulk}}$.
In addition, it would be possible to choose a semiconductor
with a large $g$-factor ($g > 2$)~\cite{Shabani2016two-dimensional,kjaergaard2016quantized,moehle2021insbas,Phan2022BFS,Reeg2018Metallization}, which
enables a larger Zeeman potential in a semiconductor $V$ than that in the
bulk of superconductor $V_{\mathrm{bulk}}$.
For example, $g \sim 10$ or larger was reported in a two dimensional Al-InAs hybrid 
system~\cite{Shabani2016two-dimensional,kjaergaard2016quantized,moehle2021insbas,Phan2022BFS}.
A condition for the appearance of the BFSs in a thin semiconductor 
\begin{align} \label{eq:condition1}
  \Delta < V = \frac{1}{2} g \mu_B B ,
\end{align}
and that for the stable superconducting state in the bulk
\begin{align} \label{eq:condition2}
  \Delta_{\mathrm{bulk}} \gg V_{\mathrm{bulk}} \sim \mu_B B ,
\end{align}
can be satisfied simultaneously 
in the proximity structure in Fig.~\ref{fig:system}.
In what follows, we also assume that the transition to superconducting phase is always the second-order~\cite{sato:prb2024}
and that the dependence of $\Delta$ on temperatures is described by BCS theory.

In a junction shown in Fig.~\ref{fig:system},
the thermoelectric current flows only in the semiconducting segment.
The electric current is absent in an insulator.
Furthermore, the thermoelectric coefficient
in the parent superconductor $  \alpha_{\mathrm{S}} \approx \alpha_{\mathrm{N}} \exp (- \Delta_{\mathrm{bulk}} / T) \ll \alpha_{\mathrm{N}}$
is almost zero at low temperatures~\cite{galperin1974thermoelectric}
with $\alpha_{\mathrm{N}}$ being the thermoelectric coefficient in the normal state.
Therefore, the electric current flowing through the semiconducting film dominates 
the thermoelectric effect of whole structure in Fig.~\ref{fig:system}.

\section{Current formula}\label{sec:formula}
The electric current in the weak coupling limit~\cite{Graf1996Electtic_thermal} is represented by
\begin{align} \label{eq:current}
  \bm{j(R)} 
  =
  \frac{e}{4} 
  \int \frac{d \bm{k} d \epsilon}{(2 \pi)^{d+1}}
  i \bm{v}
  \mathrm{Tr}
  \left[
    \check{G}^{K} (\bm{R}, \bm{k}, \epsilon)
  \right], 
\end{align}
with $\bm{v}=\hbar \bm{k}/m$ and $\bm{R}$ being the velocity and the place in
real space, respectively.
The Keldysh Green's function $\check{G}^{K}$ is the solution of the Gor'kov
equation,
\begin{align}
  &\left[
    \begin{pmatrix}
      \check{L}_0 + \check{L}_1
      & 0 \\
      0 &
      \check{L}_0 + \check{L}_1
    \end{pmatrix}
    -
    \begin{pmatrix}
      \check{\Sigma}^{R} & \check{\Sigma}^{K} \\
      0 & \check{\Sigma}^{A}
    \end{pmatrix}
  \right]_{(\bm{k,\epsilon})} \nonumber\\
 & \times  
  \begin{bmatrix}  
    \check{G}^{R} & \check{G}^{K} \\
    0 & \check{G}^{A}
  \end{bmatrix}_{(\bm{R}, \bm{k}, \epsilon)}
  =
  \begin{bmatrix}
    \check{1} & 0 \\
    0 & \check{1}
  \end{bmatrix}, \label{eq:gorkov}
\end{align}
where
\begin{align}
  \check{L}_0 (\bm{k}, \epsilon)
  &=
  \epsilon - \xi_{\bm{k}} \check{\tau}_3 - \check{V}, \\
  \check{L}_1 (\bm{k}, \epsilon) & = \frac{i}{2} \hbar \bm{v} \cdot \nabla_{\bm{R}} \check{\tau}_3, \\
  \check{\Sigma}^{X}
  &=
  \check{\Delta} - \check{\Sigma}_{\mathrm{imp}}^{X} ,
\end{align}
with
\begin{align}
  \check{\Delta}
  =&
  \begin{bmatrix}
    0 & \hat{\Delta} (\bm{k}) \\
    - \undertilde{\hat{\Delta}} (\bm{k}) & 0
  \end{bmatrix} , \\ 
  \check{V}
  =&
  \begin{bmatrix}
    \bm{V} \cdot \hat{\bm{\sigma}} + \bm{\lambda} \times \bm{k} \cdot \hat{\bm{\sigma}} &
    0 \\
    0 &
    - \bm{V} \cdot \hat{\bm{\sigma}}^{\ast} + \bm{\lambda} \times \bm{k} \cdot \hat{\bm{\sigma}}^{\ast}
  \end{bmatrix} .
\end{align}
Here $\check{\tau}_{i} (i=1,2,3)$ are Pauli's matrix in particle-hole space.
The relation 
  $\undertilde{Y} (\bm{k}, \epsilon) 
  \equiv
  Y^{\ast} (- \bm{k}, - \epsilon)$
represents particle-hole transformation of a function $Y(\bm{k}, \epsilon) $.
The self-energy due to random impurity scatterings is denoted by $\check{\Sigma}^{X}_{\mathrm{imp}}$. 
The details of the derivation is given in Appendix~\ref{AppA}.

The thermoelectric coefficient is calculated within the linear response 
to the thermal gradient which is considered through the distribution function 
\begin{align}
  \Phi (\epsilon, \bm{R})
  =
  \tanh 
  \left[
    \frac{\epsilon}{2 T(\bm{R})}
  \right] .
\end{align}
As discussed in Appendix~\ref{AppA},
the Green's function $\check{G}^{R, A}$ remains unchanged from its expression in equilibrium $\check{G}^{R, A}_0$
within the first order of $\nabla  \Phi$.
The Gor'kov equation for the Keldysh component becomes
\begin{align}
  &\left[
    \check{L}_0 (\bm{k}, \epsilon)
    -
    \check{\Sigma}^{R} (\bm{k}, \epsilon)
    +
    \frac{i}{2} \hbar \bm{v} \cdot \nabla_{\bm{R}} \check{\tau}_3
  \right]
  \check{G}^{K} (\bm{R}, \bm{k}, \epsilon) \nonumber\\
&  -
  \check{\Sigma}^{K} (\bm{k}, \epsilon)
  \check{G}_{0}^{A} (\bm{k}, \epsilon)
  = 0. \label{eq:gorkov_K}
\end{align}
We seek the solution of the Keldysh Green's function of the 
form $\check{G}^K = \check{G}^K_0 + \delta \check{G}^K$
with Eqs.~\eqref{eq:GK} and \eqref{eq:SelfK},
where the first term is the solution in equilibrium 
and the second term is the deviation from it.
In the limit of weak impurity scatterings, the solution within the first order of $\nabla \Phi$ is given by
\begin{align} 
  \delta \check{G}^{K}
  =&
  - \frac{i}{2} \hbar \bm{v} \cdot \nabla_{\bm{R}} \Phi (\epsilon, \bm{R})
  \nonumber\\
  &\times 
  \left[
    \check{G}_{0}^{R} (\bm{k}, \epsilon)
    \check{\tau}_3   \left\{
      \check{G}_{0}^{R} (\bm{k}, \epsilon)
      -
      \check{G}_{0}^{A} (\bm{k}, \epsilon)
    \right\}
  \right] . \label{eq:dgk}
\end{align}
The current density and the thermoelectric coefficient are calculated as
\begin{align}
  \bm{j}^{\mu} (\bm{R}) = - \alpha^{\mu, \nu} \nabla_{\bm{R}}^{\nu} T ,
\end{align}
where $\mu$ and $\nu$ represent the direction of the current and the temperature gradient.
Eq.~\eqref{eq:alpha1} shows the general expression of the coefficient.
Since all the off-diagonal elements of the coefficients are zero in a 
superconductor under consideration, we show 
only the diagonal elements as 
\begin{align}
  \alpha^{\nu}
  &=
  \frac{e \hbar}{32 \pi T^2}
  \int_{-\infty}^{\infty} d \epsilon
  \frac{\epsilon}{\cosh^2 (\epsilon/2T)} \nonumber\\
  \times &  \int \frac{d \hat{\bm{k}}}{2\pi} \left(\hat{k}^{\nu}\right)^2
  \int_{-\infty}^{\infty} d \xi N(\xi) v^2 (\xi) 
  I(\bm{k}, \epsilon) , \label{eq:alpha} \\
  I(\bm{k}, \epsilon)
  &= 
  \mathrm{Tr}
  \left[
    \hat{G} (\hat{G} - \hat{G}^{\dagger})
    -
    \undertilde{\hat{G}}(\undertilde{\hat{G}} - \undertilde{\hat{G}}^{\dagger}) \right.
	\nonumber\\
   & +\left.   
    \hat{F} \hat{F}^{\dagger}
    -
    \undertilde{\hat{F}} \undertilde{\hat{F}}^{\dagger}
  \right]^{R}_{(\bm{k}, \epsilon)}, \label{eq:idef1}
\end{align}
with $\nu = x$ or $y$, where $R$ in the superscript of Eq.~\eqref{eq:idef1} means that 
all the Green's functions belong to the retarded causality.
Hereafter we omit ``0" from the subscript of the Green's functions because 
the coefficient is expressed only by the Green's functions in equilibrium. 
We have used a relation
\begin{align}
\frac{1}{V_{\mathrm{vol}}} \sum_{\bm{k}} 
\to \int \frac{d \hat{\bm{k}}}{2\pi}   \int_{-\infty}^{\infty} d \xi\,  N(\xi), 
\end{align} 
with $\hat{\bm{k}}=(k_x, k_y)/|\bm{k}|$. 
The $2 \times 2$ retarded Green's functions in Eq.~\eqref{eq:idef1} are the solution of the Gor'kov equation,
\begin{align}
\left[ \epsilon + i \delta - \check{H}_{\mathrm{BdG}} \right]
\left[ \begin{array}{cc}
\hat{G} & \hat{F} \\ - \undertilde{\hat{F}} & - \undertilde{\hat{G}} 
\end{array}\right]_{(\bm{k}, \epsilon)}^{R} = \check{1}.
\end{align}
Here we introduce the lifetime $\tau=\hbar/\delta$ of superconducting states.
The Green's functions for the BdG Hamiltonian in Eq.~\eqref{eq:BdG} 
are calculated to be
\begin{align}
\hat{G}(\bm{k}, \epsilon) =& \frac{1}{Z}\left[
(\epsilon - \xi_{\bm{k}}) \, \undertilde{z}_{\mathrm{N}} 
- (\epsilon+\xi_{\bm{k}}) \Delta^2 \right. \nonumber\\
& \left.
-\bm{V} \cdot \hat{\bm{\sigma}} ( \undertilde{z}_{\mathrm{N}} +\Delta^2) 
- \bm{\alpha}_{\bm{k}} \cdot \hat{\bm{\sigma}}
 ( \undertilde{z}_{\mathrm{N}} -\Delta^2)
\right], \label{eq:gba1}\\
\hat{F}(\bm{k}, \epsilon) =& \frac{1}{Z}\left[ (\epsilon^2 - \xi_{\bm{k}}^2 -\Delta^2 -\bm{\alpha}_{\bm{k}}^2 + 
\bm{V}^2) - 2 \epsilon  \bm{V} \cdot \hat{\bm{\sigma}} 
\right. \nonumber\\
&\left.
- 2\xi_{\bm{k}} \bm{\alpha}_{\bm{k}} \cdot \hat{\bm{\sigma}} 
- 2 i \bm{V} \times \bm{\alpha}_{\bm{k}} \cdot \hat{\bm{\sigma}} 
\right] \Delta i \hat{\sigma}_y,\label{eq:fba1}\\
Z= & (\epsilon^2 - \xi_{\bm{k}}^2 -\Delta^2 -\bm{\alpha}_{\bm{k}}^2 
+ \bm{V}^2)^2 
+ 4  (\bm{V} \times \bm{\alpha}_{\bm{k}})^2 \nonumber\\
& 
- 4 (\epsilon \,\bm{V} + \xi_{\bm{k}}
 \bm{\alpha}_{\bm{k}} )^2,\label{eq:zba1}\\
\undertilde{z}_{\mathrm{N}}=& (\epsilon+\xi_{\bm{k}})^2 - ( \bm{V} - \bm{\alpha}_{\bm{k}} )^2, \quad
\bm{\alpha}_{\bm{k}} = \bm{\lambda} \times \bm{k}.
\end{align}
The density of states per volume are calculated as
\begin{align}
N(E)
&=  \int \frac{d \hat{\bm{k}}}{2\pi} \, n(\hat{\bm{k}}, E), \label{eq:dos}\\
n(\hat{\bm{k}}, E) &= \frac{-N_0}{8\pi i}
\int_{-\infty}^{\infty} d \xi \,
\mathrm{Tr}
\left[
   \hat{G} - \hat{G}^{\dagger}
  - \undertilde{\hat{G}} + \undertilde{\hat{G}}^{\dagger} \right]^{R}_{(\bm{k}, E)},
\label{eq:ardos}
\end{align}
where $n(\hat{\bm{k}}, E)$ is the angle-resolved density of states
and $N_0$ is the DOS per volume in the normal states at the Fermi level.

For numerical simulation, we put $\delta = 10^{-4}  \Delta_0 $ and $\Delta_0 = 0.02 \mu$, 
where $\Delta_0$ is the amplitude of induced pair potential at zero temperature.

\section{Results}\label{Sec3}

\subsection{Density of statets}
\begin{figure}[ttt]
  \includegraphics[width=8cm]{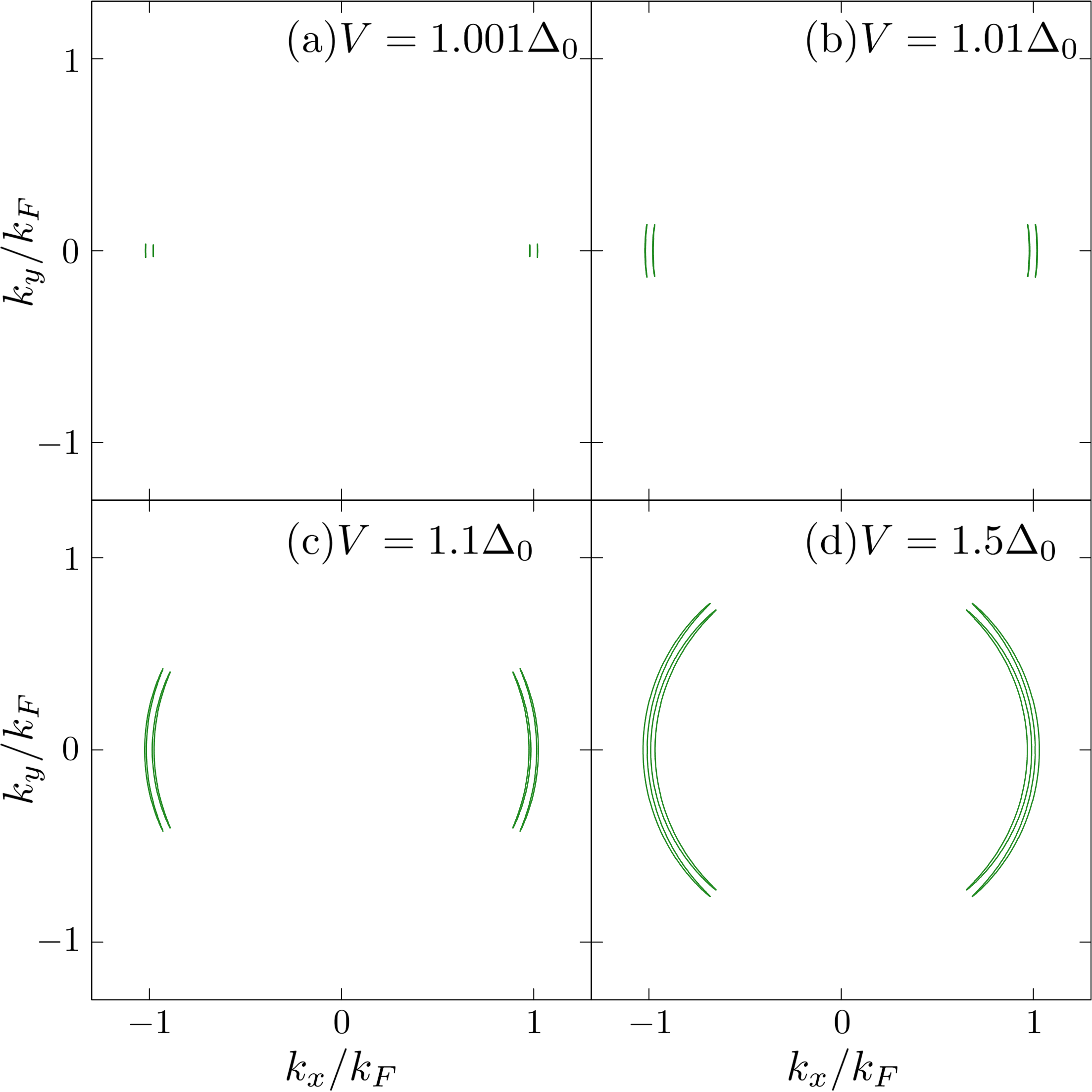}
  \caption{
    The BFSs for $V = 1.001\Delta_0$, $V = 1.01\Delta_0$, 
    $V = 1.1\Delta_0$, and $V = 1.5\Delta_0$ are displayed in (a), (b), (c), and (d), respectively.
    The strength of the Rashba SOI is fixed at $\lambda k_F = 2 \Delta_0$.
    }
  \label{fig:BFS}
\end{figure}
In Fig.~\ref{fig:BFS}, we plot the BFSs (possible wavenumber for zero-energy eigenvalue) 
for several choices of Zeeman potentials $V= \frac{1}{2}g \, \mu_B\, B$.
The size of the BFSs increases with the increase of $V$.
For Zeeman potentials slightly larger than $\Delta_0$ in Figs.~\ref{fig:BFS}(a) and (b), 
the quasiparticle states at zero energy are present around 
$\bm{k}=(k_x, k_y)=(\pm k_F, 0)$ and absent around $\bm{k}=(0, \pm k_F)$.
The anisotropy of the BFSs is derived from the anisotropy in the normal states
described by Eq.~\eqref{eq:hn}. For $\bm{V}=0$ and $\bm{\lambda}=0$,  
the original Fermi surface given by $k_x^2+k_y^2=k_F^2$ has a circular shape 
in momentum space.
The isotropy of the Fermi surface is preserved for $\bm{V}\neq 0$ and $\bm{\lambda}=0$
because the direction of a Zeeman field does not couple to any directions in the momentum space.
A Zeeman field splits the doubly degenerate Fermi surface into two.
The isotropy is preserved also for $\bm{V}= 0$ and $\bm{\lambda}\neq 0$
because the Rashba SOI preserves rotation symmetry along $z$ axis.
The Rashba SOI also splits the doubly degenerate Fermi surface into two.
In the presence of the two interactions simultaneously  
$\bm{V}\neq 0$ and $\bm{\lambda}\neq 0$, the SOI couples momentum space to spin space 
and a Zeeman field specifies a special direction in momentum space. 
Thus, the anisotropy in the Fermi surface in the normal state is a result of the coexistence 
of the Rashba SOI and the Zeeman field.
The BFSs in Fig.~\ref{fig:BFS} inherit such the anisotropy even in the superconducting 
state. The BFSs always appear in the directions perpendicular to a Zeeman field $\bm{V}$.
The results in Fig.~\ref{fig:BFS} show the BFSs aronund $\boldsymbol{k} =(k_x, k_y) \approx ( \pm k_F, 0)$ 
because a Zeeman field is applied in the $y$ direction. 
The anisotropy of the BFSs is a source of the anisotropy in the thermoelectric effect.

In Fig.~\ref{fig:BFS},
the strength of the Rashba SOI is fixed at $\lambda k_F = 2 \Delta_0$.
Changing the amplitude of $\lambda$ shifts the place of the BFSs only slightly in momentum space.
We choose $\lambda k_F = 2 \Delta_0$ throughout this paper because
the characteristic features of the BFSs shown in Fig.~\ref{fig:BFS} are insensitive 
to the choice of $\lambda$.
A role of SOI in the formation of BFSs are discussed in Appendix~\ref{AppB}.

In Fig.~\ref{fig:dos}, we present the DOS in the presence of the BFSs for several Zeeman potentials.
The DOS at zero energy remains a finite value for all the Zeeman potentials.
At $V=0$, the DOS has two coherence peaks at $E=\pm \Delta$. 
Zeeman potentials shift these peaks depending on spin of an electron, which
explains multiple peaks in $E/\Delta_0 = 2$ in Fig.~\ref{fig:dos}.
As Zeeman potentials increase, the width of a plateau near zero energy increases.
This makes a superconducting phase with the BFSs unstable thermodynamically.
To gain the condensation energy, the gapped energy spectra in DOS are necessary. 
The two-dimensional superconducting states with the BFSs are stabilized by the 
superconducting condensate in the parent superconductor.

\begin{figure}[ttt]
  \includegraphics[width=8cm]{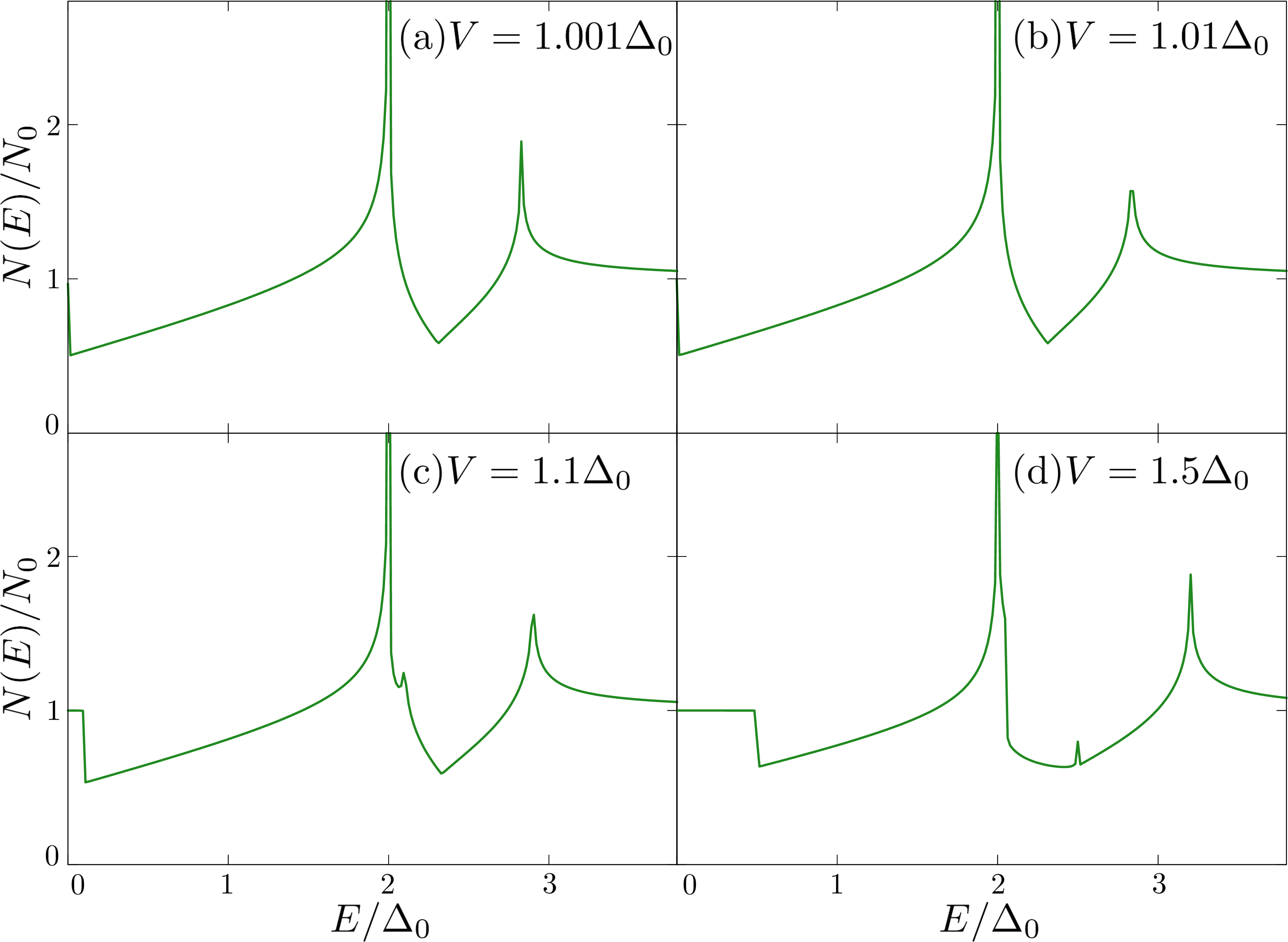}
  \caption{Density of states calculated for several Zeeman potentials at $\lambda k_F = 2 \Delta_0$.}
  \label{fig:dos}
\end{figure}

\subsection{Thermoelectric effect} 
\begin{figure}[ttt]
  \includegraphics[width=8cm]{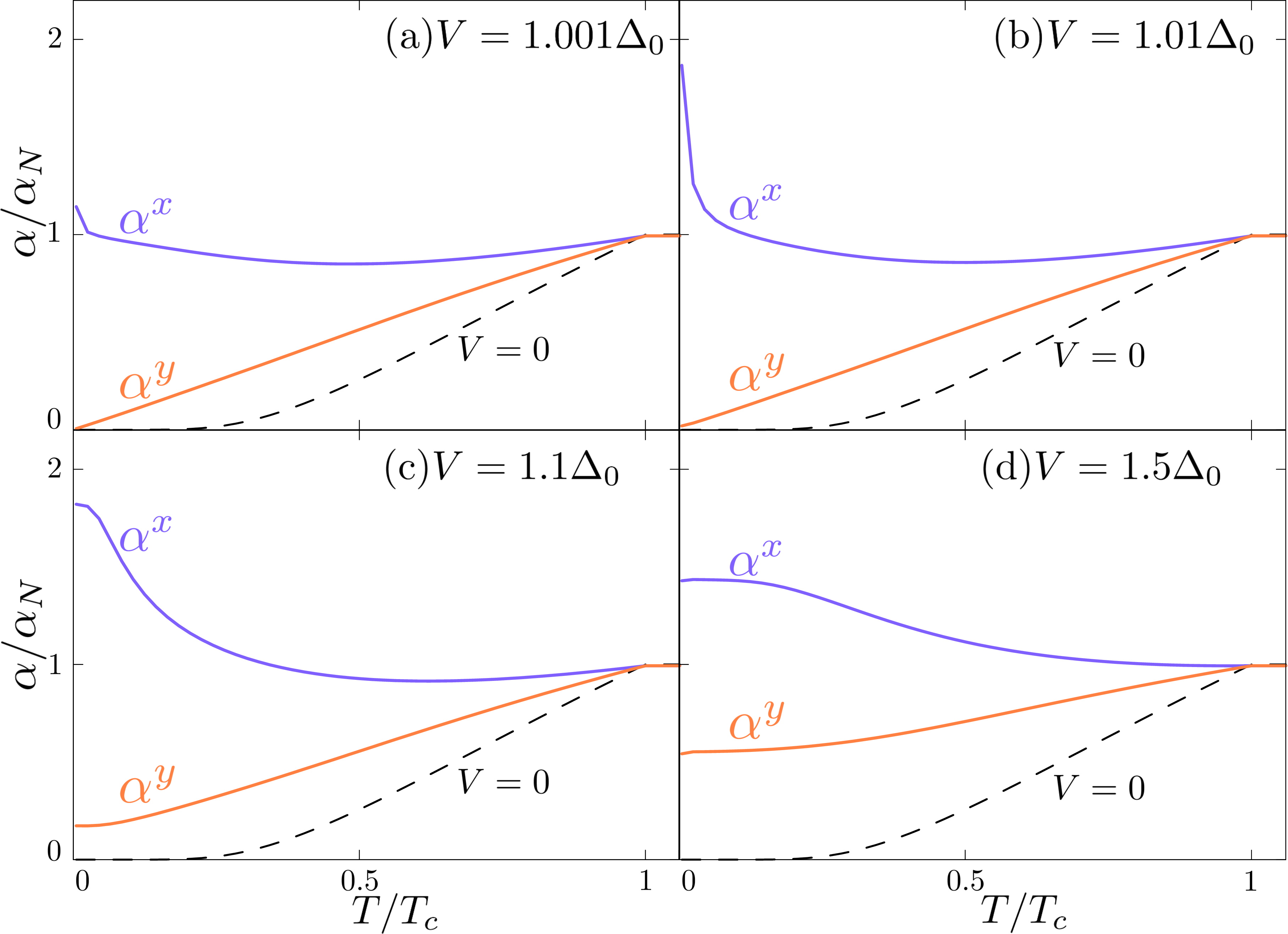}
  \caption{The thermoelectric coefficient versus temperature.
  The coefficients in the two directions are shown with $\alpha^x$ and $\alpha^y$.
  The broken lines show the results in the absence of a Zeeman potential.
  }
  \label{fig:thermoele}
\end{figure}
The calculated results of the thermoelectric coefficients 
in the presence of the BFSs are plotted as a function of temperature for several of Zeeman 
potentials in Fig~\ref{fig:thermoele}, where the dependence of $\Delta$ on temperatures 
is described by BCS theory. 
The vertical axis is normalized to the coefficient in the normal state $\alpha_{\mathrm{N}}$ 
which is isotropic in real space.
The results in the absence of a Zeeman potential are shown with a broken line 
for comparison and
obey $\alpha \approx \alpha_{\mathrm{N}}\, \exp( -\Delta / T)$ at low temperatures. 
Solid lines represent $\alpha^{x(y)}$ in which the electric current 
is perpendicular to magnetic field 
$ \bm{j} \perp \bm{B}$ ( parallel to magnetic field 
$\bm{j} \parallel \bm{B}$).
The thermoelectric coefficients indicate two characteristic features: 
the remarkable anisotropy in real space 
and $\alpha^x > \alpha_{\mathrm{N}}$.
The anisotropy in the thermoelectric effect originates from the anisotropy of the 
BFSs in momentum space shown in Fig.~\ref{fig:BFS}.
The zero-energy states around $\bm{k}=( \pm k_F,0)$ 
in Fig.~\ref{fig:BFS} can carry the electric 
current in the $x$ direction.
However, zero-energy states are absent around $\bm{k}=(0, \pm k_F)$, which results 
in the monotonic decrease of $\alpha^y$ with decreasing temperatures in Fig~\ref{fig:thermoele}.
Such anisotropy is absent in a $d$-wave superconductor with the BFSs 
(See Appendix~\ref{AppC} for details).

\begin{figure}[ttt]
  \includegraphics[width=8cm]{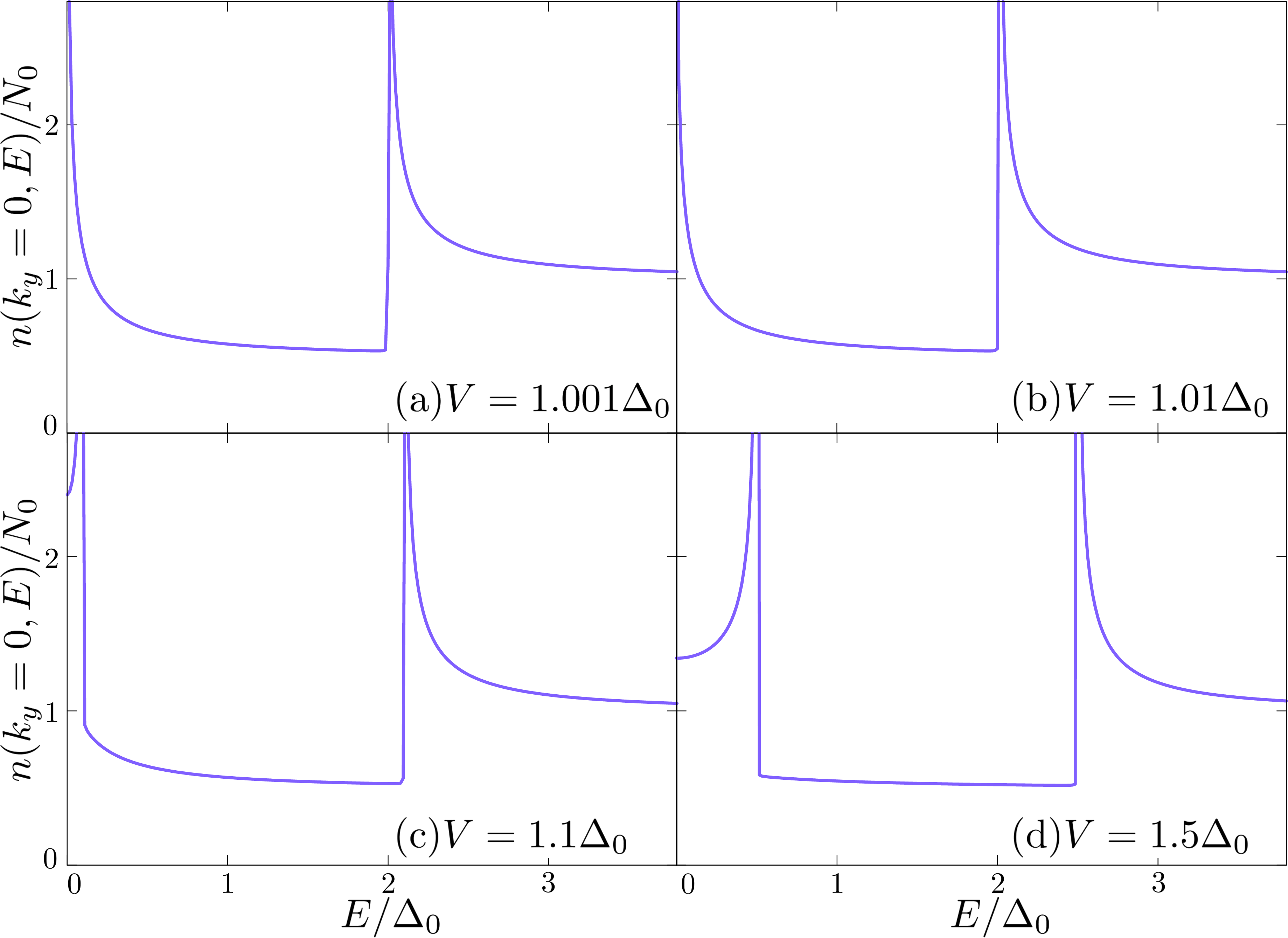}
  \caption{
    The angle-resolved density of states $n(\hat{\bm{k}}, E)$ at $k_y = 0$ 
    are shown for several Zeeman fields with $\lambda k_F = 2 \Delta_0$.
  } 
  \label{fig:dos_x}
  \includegraphics[width=8cm]{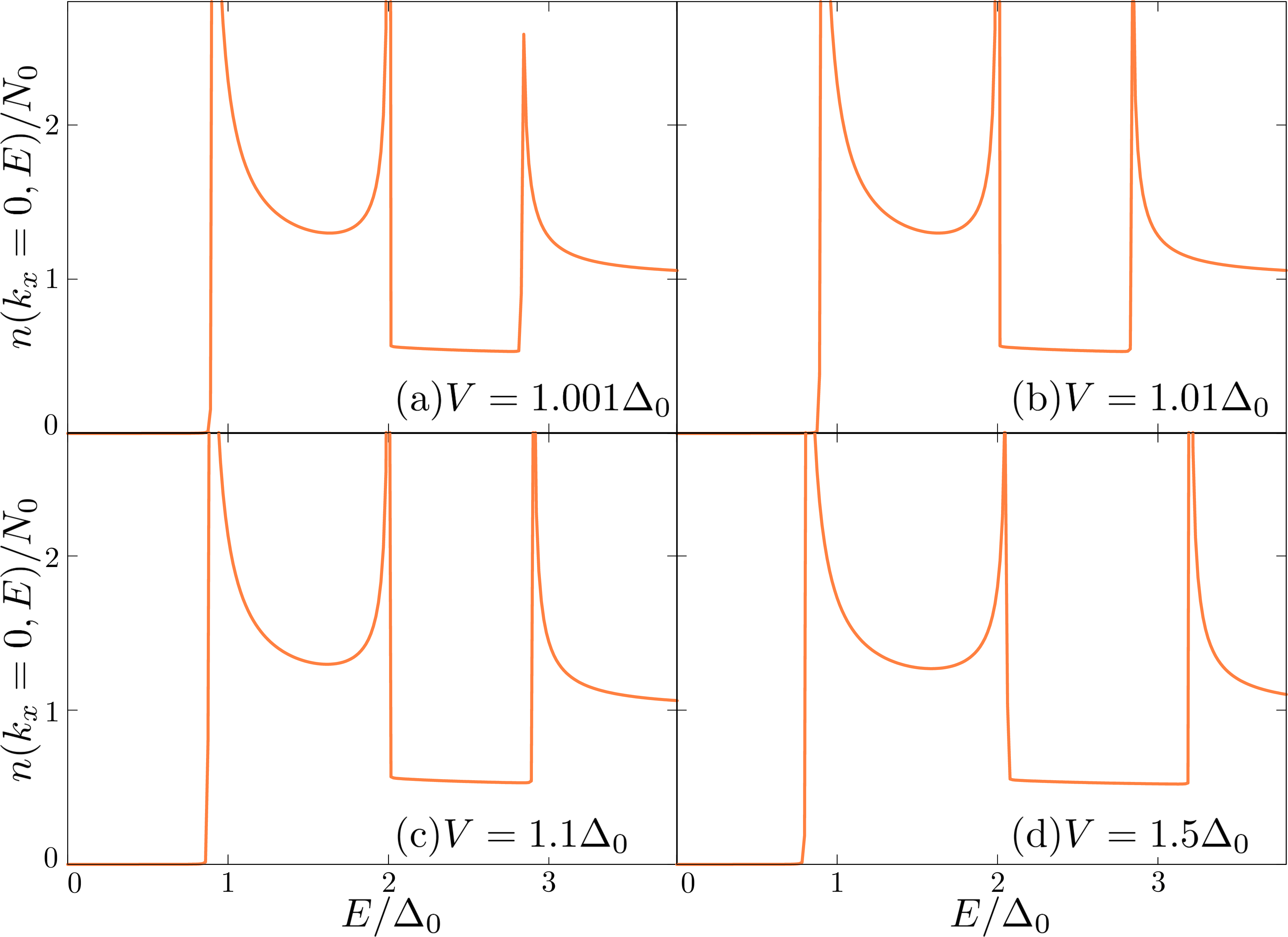}
  \caption{
    The angle-resolved density of states $n(\hat{\bm{k}}, E)$ at $k_x = 0$ 
    are shown for several Zeeman fields with $\lambda k_F = 2 \Delta_0$.
  }
  \label{fig:dos_y}
\end{figure}

To understand the unusual dependence of $\alpha^x$ on temperature, we 
analyze the angle-resolved DOS displayed in Fig.~\ref{fig:dos_x}, where
$n(\hat{\bm{k}}, E)$ along the $k_x$ axis is shown by fixing $k_y$ at 0. 
The angle-resolved DOS has two peaks for $E>0$ because a Zeeman potential shifts the coherence peaks 
depending on spins of a quasiparticle. 
For $V \approx \Delta$ in Figs.~\ref{fig:dos_x}(a) and (b),
a peak appears almost zero energy in the angle-resolved DOS. 
A quasiparticle at such zero-energy states carries the thermoelectric current 
in the $x$ direction. 
The relation $\alpha^x > \alpha_{\mathrm{N}}$ is a result of the shift 
of the coherence peak by a Zeeman potential.
The DOS along the $k_y$ axis calculated with putting $k_x=0$ is shown in Fig.~\ref{fig:dos_y}.
In contrast to Fig.~\ref{fig:dos_x}, the angle resolved DOS 
always has the gapped energy spectra at zero energy as a result of 
the absence of the BFSs around $\bm{k}=(0, \pm k_F)$ in Fig.~\ref{fig:BFS}.
This also explains the monotonic dependence of $\alpha^y$ on temperatures in Fig.~\ref{fig:thermoele}. 
The BFSs spread in the $k_y$ direction with increasing Zeeman potential as shown in Fig.~\ref{fig:BFS}.
A quasiparticle on the BFS for $k_y \neq 0$ has a finite velocity in the $y$ direction. 
Such quasiparticle states can carry the thermoelectric current in the $y$ direction. 
In Fig.~2, the number of quasiparticle states for $k_y \neq 0$ increases with increasing $V$. 
As a result, $\alpha^y$ takes a finite value and increases even at zero temperature with increasing $V$.
We conclude that the thermoelectric effect in a junction in Fig.~\ref{fig:system} 
is remarkably anisotropic in real space because of the anisotropy of the BFSs in momentum
space. 
The thermoelectric coefficient for the current 
perpendicular to a magnetic field can be larger than that in the normal state 
because a Zeeman potential shifts the gapped spectra in DOS depending on spin 
of a quasiparticle.

\section{Discussion}\label{Sec4}
Two of authors have shown that odd-frequency Cooper pairs 
coexist with quasiparticles on the BFSs~\cite{Kim2021BFSandodd-w}.
This can be seen in the expression of the anomalous Green's function in Eq.~\eqref{eq:fba1}.
The first term represents a spin-singlet Cooper pairs and
is linked to the pair potential through the gap equation.
The other terms represent spin-triplet Cooper pairs generated by a Zeepan potential and/or the SOI.
To make discussion simpler, we put $\bm{V}=\bm{\alpha}_{\bm{k}}=0$ at the denominator
of the Green function in Eq.~\eqref{eq:zba1}. 
By applying the analytic continuation $\epsilon + i\delta \to i \omega_n$ in such a situation,
the second term in Eq.~\eqref{eq:fba1} is relating to odd-frequency Cooper pairs because 
it is an odd function of the Matsubara frequency $\omega_n$.
Before calculating the thermoelectric coefficients, 
we had expected the contributions of odd-frequency Cooper pairs to the electric current.
Therefore, we formulate the thermoelectric coefficients in terms of the 
Green's functions of the Gor'kov equation.
In fact, the expression in Eq.~\eqref{eq:idef1} includes the anomalous 
Green's function $\hat{F}$ and $\undertilde{\hat{F}}$.
Unfortunately, however, we find that 
$\mathrm{Tr} [ \hat{F} \hat{F}^\dagger - \undertilde{\hat{F}} \undertilde{\hat{F}}^\dagger ] = 0$
for any Zeeman fields and SOIs in a spin-singlet $s$-wave superconductor.
Only the normal Green's functions $\hat{G}$ and $\undertilde{\hat{G}}$ contribute to the thermoelectric 
coefficient in our model.
As a result, the characteristic behavior of the thermoelectric coefficients 
can be explained well by the residual DOS due to quasiparticles on the BFSs.

\section{Conclusion}\label{Sec5}
We have studied the thermoelectric effect in a thin superconducting film 
between a conventional superconductor and an insulator.
In the presence of an external magnetic field and the Rashba spin-orbit interactions,
the superconducting phase having the Bogoliubov-Fermi surfaces can be realized 
at the thin film. 
The thermoelectric coefficient is calculated based on the linear response
theory in the presence of the spatial gradient of a temperature. 
The calculated results of the thermoelectric coefficients show the remarkable
anisotropy at low temperatures: the coefficient for the current perpendicular to
a magnetic field is larger than that for the current parallel to a magnetic field. 
Moreover, the coefficient for the current perpendicular to
a magnetic field can be larger than its normal state value. 
These characteristic features of thermoelectric effect are explained well 
by the anisotropy of the Bogoliubov-Fermi surfaces in momentum space.
Our results indicate a way of realizing the stable Bogoliubov-Fermi surfaces and
how to check the existence of the Bogoliubov-Fermi surfaces.

\section*{Acknowledgments}
The authors are grateful to Y.~Tanaka, Y.~Kawaguchi, and S.~Hoshino for useful discussions.
T. Sato was supported by JST, the establishment of university fellowships towards the creation of science technology innovation, Grant Number JPMJFS2101.
S.I. is supported by the Grant-in-Aid for Early-Career Scientists (JSPS KAKENHI Grant No. JP24K17010).
S. K. was supported by JSPS KAKENHI (Grants No. JP19K14612 and No. JP22K03478) and JST 
CREST (Grant No. JPMJCR19T2).

\appendix

\begin{widetext}
\section{Formalism}\label{AppA}
\subsection{Gor'kov equation}
We begin with the general expression of the electric current in terms of Keldysh Green's funcion,
\begin{align}
  G_{\alpha, \beta}^{K} (\mathrm{x_1, x_2})
  &=
  -i 
  \big\langle
    \psi_{\alpha}(\mathrm{x_1}) \psi_{\beta}^{\dagger}(\mathrm{x_2})
    -
    \psi_{\beta}^{\dagger}(\mathrm{x_2}) \psi_{\alpha}(\mathrm{x_1})
  \big\rangle , \\
  \underline{G}_{\alpha, \beta}^{K} (\mathrm{x_1, x_2})
  &=
  -i 
\big\langle
    \psi_{\alpha}^{\dagger}(\mathrm{x_1}) \psi_{\beta}(\mathrm{x_2})
    -
    \psi_{\beta}(\mathrm{x_2}) \psi_{\alpha}^{\dagger}(\mathrm{x_1})
  \big\rangle ,
\end{align}
where $\mathrm{x}=(\bm{r},t)$ is the combined representation of coordinate 
and $\psi_{\alpha}(\mathrm{x})$ ($\psi_{\alpha}^{\dagger}(\mathrm{x})$) is the annihilation(creation) operator 
of an electron.
By using the anticommutation relations 
and applying $\nabla_{\bm{r}_1}$ from the left,
the current density is expressed in terms of the Keldysh Green's function
\begin{align}
  \bm{j} (\mathrm{x}_1)
  =
  \frac{e \hbar}{4 m}
  \lim_{\mathrm{x_1} \rightarrow \mathrm{x_2}}
  \nabla_{\bm{r}_1} \mathrm{Tr} 
  \left[
    \check{G}^{K} (\mathrm{x_1, x_2})
  \right] , \label{eq:a_current}
\end{align}
where $-e$ is the charge of an electron,
$\mathrm{Tr}$ is the trace in spin space and particle-hole space
\begin{align}
  \mathrm{Tr}
  \left[
    \check{G}^{K}(\mathrm{x_1, x_2})
  \right]
  =
  \sum_{\alpha}
  \left(
    G_{\alpha, \alpha}^{K} + \underline{G}_{\alpha, \alpha}^{K}
  \right) .
\end{align}

The Green's function is a solution of the Gor'kov equation which describes
the superconducting states 
\begin{align}
  \int d \mathrm{x_2}
  \begin{bmatrix}
    \check{L} - \check{\Sigma}^{R} & - \check{\Sigma^{K}} \\
    0 & \check{L} - \check{\Sigma}^{A}
  \end{bmatrix}_{(\mathrm{x_1, x_2})}
  \begin{bmatrix}
    \check{G}^{R} & \check{G}^{K} \\
    0 & \check{G}^{A}
  \end{bmatrix}_{(\mathrm{x_2, x_3})}
  =
  \begin{bmatrix}
    \check{1} & 0 \\
    0 & \check{1}
  \end{bmatrix}
  \delta(\mathrm{x_1 - x_3}) , \label{eq:gorkov1_a}\\
  \check{L} (\mathrm{x_1, x_2})
  =
  i \delta(\bm{r}_1 - \bm{r}_2) \partial_{t_2}
  - \delta(t_1 - t_2)
  \left\{
    - \frac{\hbar^2}{2 m} \nabla_{\bm{r}_2}^{2} - \mu
  \right\} \check{\tau}_3
  -
  \check{V} (\mathrm{x_1, x_2}), \\
  \check{V} (\mathrm{x_1, x_2})
  =
  \delta(\mathrm{x_1 - x_2})
  \begin{bmatrix}
    \bm{V} \cdot \hat{\bm{\sigma}} 
    - i \bm{\lambda} \times \hat{\nabla}_{\bm{r}_2} \cdot \hat{\bm{\sigma}} 
    & 0 \\
    0 &
    - \bm{V} \cdot \hat{\bm{\sigma}}^{\ast} 
    - i \bm{\lambda} \times \hat{\nabla}_{\bm{r}_2} \cdot \hat{\bm{\sigma}}^{\ast}
  \end{bmatrix},
\end{align}
where $\check{\tau}_i (i=1,2,3)$ are Pauli's matrix in particle-hole space.
The self-energy consists of two contributions,
\begin{align}
  \check{\Sigma}^{X} (\mathrm{x_1, x_2})
  =
  \check{\Delta}(\mathrm{x_1, x_2})
  +
  \check{\Sigma}_{\mathrm{imp}}^{X} (\bm{r}_1 - \bm{r}_2)
  \delta(t_1 - t_2), 
  \ \ (X = R, A, K)  .
\end{align}
The pair potential 
\begin{align}
  \check{\Delta} (\mathrm{x_1, x_2})
  =
  \begin{bmatrix}
    0 & \hat{\Delta} (\mathrm{x_1 - x_2}) \\
    - \hat{\Delta}^{\ast} (\mathrm{x_1 - x_2}) & 0
  \end{bmatrix},
\end{align}
is the self-energy due to the attractive interactions.
The impurity self-energy is uniform in real space due to 
ensemble averaging and instantaneous in time.

\subsection{Mixed representation}

In what follows, we apply the mixed representation to the Green's function
\begin{align}
  \check{G} (\mathrm{x_1, x_2})
  =
  \check{G} (\bm{R}, \bm{r}, T, t) 
  =
  \int \frac{d \bm{k} d \epsilon}{(2 \pi)^{d+1}}
  \check{G} (\bm{R}, \bm{k}, T, \epsilon)
  e^{i (\bm{k} \cdot \bm{r} - \epsilon t)} ,
\end{align}
with
\begin{align}
  \bm{R} = \frac{\bm{r}_1 + \bm{r}_2}{2} , \quad
  \bm{r} = \bm{r}_1 - \bm{r}_2 , \quad
  T = \frac{t_1 + t_2}{2} , \quad
  t = t_1 - t_2 .
\end{align}
The Green's function is independent of center-of-mass-time 
$T$ because we consider time-independent phenomena in this paper.
We have applied the Fourier transformation to the internal degree of freedom
$\mathrm{x_1 - x_2}$.
Since $\nabla_{\bm{r}_1} = \nabla_{\bm{R}}/2 + \nabla_{\bm{r}}$, 
we obtain
\begin{align}
  \nabla_{\bm{r}_1} &G(\mathrm{x_1, x_2}) 
  =
  \int \frac{d \bm{k} d \epsilon}{(2 \pi)^{d+1}}
  \left[
    \frac{1}{2} \nabla_{\bm{R}} + i \bm{k}
  \right]
  G(\bm{R, k}, \epsilon)
  e^{i (\bm{k \cdot r} - \epsilon t)} .
\end{align}
In the mean-field theory of superconductivity, 
the spatial gradient of the Green's function is estimated as
$\left| \nabla_{\bm{R}}G(\bm{R, k}, \epsilon) \right| \simeq G(\bm{R, k}, \epsilon)/\xi_0$,
whereas the dominant wavenumber to the integral is 
$|\bm{k}| \simeq k_F \simeq 1/\lambda_F$.
The coherence length 
$\xi_0 = \hbar v_F/(\pi \Delta_0) \simeq \epsilon_F /(\pi k_F T_c)$
is much longer than the Fermi wavelength $\lambda_F$.
The first term is negligible because the transition temperature $T_c$
is much smaller than the Fermi energy $\epsilon_F$.
By applying the argument to 
the current density in Eq.~\eqref{eq:a_current}, we obtain 
Eq.~\eqref{eq:current} in the text.

The derivative in $\check{L}$ is carried out as
\begin{align}
  \nabla_{\bm{r}_2} \check{G} (\mathrm{x_2, x_3})
  &=
  \int \frac{d \bm{k} d \epsilon}{(2 \pi)^{d+1}}
  \left\{
    \frac{1}{2} \nabla_{\bm{R}_{23}} + \nabla_{\bm{r}_{23}}
  \right\}
  \check{G} (\bm{R}_{23}, \bm{k}, \epsilon)
  e^{i (\bm{k \cdot r}_{23} - \epsilon t_{23})}, \\
  &\simeq
  \int \frac{d \bm{k} d \epsilon}{(2 \pi)^{d+1}}
  i \bm{k} 
  \check{G} (\bm{R}_{23}, \bm{k}, \epsilon)
  e^{i (\bm{k \cdot r}_{23} - \epsilon t_{23})} , \\
  \nabla_{\bm{r}_2}^{2} \check{G} (\mathrm{x_2, x_3})
  &=
  \int \frac{d \bm{k} d \epsilon}{(2 \pi)^{d+1}}
  \left\{
    \frac{1}{4} \nabla_{\bm{R}_{23}}^{2} 
    + \nabla_{\bm{R}_{23}} \cdot \nabla_{\bm{r}_{23}}
    + \nabla_{\bm{r}_{23}}^{2}
  \right\}
  \check{G} (\bm{R}_{23}, \bm{k}, \epsilon)
  e^{i (\bm{k \cdot r}_{23} - \epsilon t_{23})}, \nonumber \\
  &\simeq
  \int \frac{d \bm{k} d \epsilon}{(2 \pi)^{d+1}}
  \left\{
    i \nabla_{\bm{R}_{23}} \cdot \bm{k} - \bm{k}^{2}
  \right\}
  \check{G} (\bm{R}_{23}, \bm{k}, \epsilon)
  e^{i (\bm{k \cdot r}_{23} - \epsilon t_{23})} , \\
  \partial_{t_{2}} \check{G} (\mathrm{x_2, x_3})
  &=
  \int \frac{d \bm{k} d \epsilon}{(2 \pi)^{d+1}}
  \left\{
    \frac{1}{2} \partial_{T_{23}} + \partial_{t_{23}}
  \right\}
  \check{G} (\bm{R}_{23}, \bm{k}, \epsilon)
  e^{i (\bm{k \cdot r}_{23} - \epsilon t_{23})}, \nonumber \\
  &=
  \int \frac{d \bm{k} d \epsilon}{(2 \pi)^{d+1}}
  i \epsilon
  \check{G} (\bm{R}_{23}, \bm{k}, \epsilon)
  e^{i (\bm{k \cdot r}_{23} - \epsilon t_{23})}.
\end{align}
By substituting these results into Eq.~\eqref{eq:gorkov1_a}, 
we obtain the Gor'kov equation for the mixed-representation 
in Eq.~\eqref{eq:gorkov}.

\subsection{Current formula in the linear response}
The spatial gradient of temperature is considered through the distribution function
\begin{align}
  \Phi (\epsilon, \bm{R})
  =
  \tanh 
  \left[
    \frac{\epsilon}{2 T(\bm{R})}
  \right] . \label{eq:dist1}
\end{align}
In what follows, we derive the current in Eq.~\eqref{eq:current} within the first 
order of $\nabla \Phi$.
The Gor'kov equation for $\check{G}^{R, A}$ becomes
\begin{align}
  \left[
    \check{L}_0 (\bm{k}, \epsilon)
    -
    \check{\Sigma}^{R, A} (\bm{k}, \epsilon)
    +
    \frac{i}{2} \hbar \bm{v} \cdot \nabla_{\bm{R}} \check{\tau}_3
  \right]
  \left[
    \check{G}_{0}^{R, A} (\bm{k}, \epsilon)
    +
    \delta \check{G}^{R, A} (\bm{R}, \bm{k}, \epsilon)
  \right]
  = \check{1},
\end{align}
where $\delta \check{G}^{R, A}$ is the deviation of the Green's function
from their values in equilibrium $\check{G}^{R, A}_0$.
Within the first order, we obtain
\begin{align}
  \left[
    \check{L}_0 (\bm{k}, \epsilon)
    -
    \check{\Sigma}^{R, A} (\bm{k}, \epsilon)
  \right]
  \delta \check{G}^{R, A} (\bm{R}, \bm{k}, \epsilon) 
  = 0,
\end{align}
because $\nabla_{\bm{R}} \check{G}^{R, A}_0 = 0$ and Gor'kov equation in equilibrium
$
  \left[
    \check{L}_0 (\bm{k}, \epsilon)
    -
    \check{\Sigma}^{R, A} (\bm{k}, \epsilon)
  \right]
  \check{G}^{R, A}_0 (\bm{k}, \epsilon)
  = \check{1}
$.
The solution is $\delta \check{G}^{R, A} (\bm{R}, \bm{k}, \epsilon) = 0$.
The Gor'kov equation for the Keldysh component is given in Eq.~\eqref{eq:gorkov_K}
We seek the solution of the form 
\begin{align}
  \check{G}^{K} (\bm{R}, \bm{k}, \epsilon)
  &=
  \check{G}_{0}^{R} (\bm{k}, \epsilon) \Phi(\epsilon, \bm{R})
  -
  \Phi(\epsilon, \bm{R}) \check{G}_{0}^{A} (\bm{k}, \epsilon)
  +
  \delta \check{G}^{K} (\bm{R}, \bm{k}, \epsilon) , \label{eq:GK} \\
  \check{\Sigma}^{K} (\bm{k}, \epsilon)
  &=
  \check{\Sigma}^{R} (\bm{k}, \epsilon) \Phi(\epsilon, \bm{R})
  - 
  \Phi(\epsilon, \bm{R}) \check{\Sigma}^{A} (\bm{k}, \epsilon)
  +
  \delta \check{\Sigma}^{K} (\bm{R}, \bm{k}, \epsilon). \label{eq:SelfK}
\end{align}
The first term and second term in Eq.~\eqref{eq:GK} and ~\eqref{eq:SelfK} are the solutions in equilibrium.
We put $\delta \check{\Sigma}^{K} = 0$ because the deviation of the self-energy
$\check{\Sigma}$ is considered through the distribution function.
The solution is presented in Eq.~\eqref{eq:dgk}.
By substituting the results into Eq.~\eqref{eq:current},
the current density is calculated to be
\begin{align}
  \bm{j}^{\mu} (\bm{R})
  &=
  \frac{e \hbar}{8}
  \int \frac{d \bm{k} d \epsilon}{(2 \pi)^{d+1}}
  \bm{v}^{\mu} \bm{v}^{\nu} \cdot \left(\nabla_{\bm{R}}^{\nu} \Phi\right)
  \mathrm{Tr}
  \left[
    \check{G}_{0}^{R} (\bm{k}, \epsilon)
    \check{\tau}_3
    \left\{
      \check{G}_{0}^{R} (\bm{k}, \epsilon)
      -
      \check{G}_{0}^{A} (\bm{k}, \epsilon)
    \right\}  
  \right] , \\
  &=
  - \alpha^{\mu, \nu} \nabla_{\bm{R}}^{\nu} T .
\end{align}
The thermoelectric coefficient is given by
\begin{align}
  \alpha^{\mu, \nu}
  &=
  \frac{e \hbar}{32 \pi T^2}
  \int_{-\infty}^{\infty} d \epsilon
  \frac{\epsilon}{\cosh^2 (\epsilon/2T)}
  \int \frac{d \bm{k}}{(2 \pi)^d} \bm{v}^{\mu} \bm{v}^{\nu}
  I(\bm{k}, \epsilon) , \\
  &=
  \frac{e \hbar}{32 \pi T^2}
  \int_{-\infty}^{\infty} d \epsilon
  \frac{\epsilon}{\cosh^2 (\epsilon/2T)}
  \int \frac{d \hat{\bm{k}}}{S_d} \hat{k}^{\mu} \hat{k}^{\nu}
  \int_{-\infty}^{\infty} d \xi N(\xi) v^2 (\xi) 
  I(\bm{k}, \epsilon) , \label{eq:alpha1}\\
  I(\bm{k}, \epsilon)
  &\equiv
  \mathrm{Tr}
  \left[
    \check{G}_{0}^{R} (\bm{k}, \epsilon)
    \check{\tau}_3
    \left\{
      \check{G}_{0}^{R} (\bm{k}, \epsilon)
      -
      \check{G}_{0}^{A} (\bm{k}, \epsilon)
    \right\}
  \right] , \\
  &= \label{Ik_spin}
  \mathrm{Tr}
  \left[
    \hat{G} (\hat{G} - \hat{G}^{\dagger})
    -
    \undertilde{\hat{G}}(\undertilde{\hat{G}} - \undertilde{\hat{G}}^{\dagger})
    +
    \hat{F} \hat{F}^{\dagger}
    -
    \undertilde{\hat{F}} \undertilde{\hat{F}}^{\dagger}
  \right]_{(\bm{k}, \epsilon)}^{R}
\end{align}
where $\mu$ and $\nu$ are respectively 
the direction of current density and temperature gradient.
$\mathrm{Tr}$ in Eq.~\eqref{Ik_spin} means the trace in spin space.
The contribution from the anomalous Green's function vanishes 
by applying $\mathrm{Tr}$ for any Zeeman fields and spin-orbit interactions in a spin-singlet s-wave superconductor.
We have used the structure of the Green's function in particle-hole space represented as
\begin{align}
  \check{G}^{R, A} (\bm{k}, \epsilon) 
  =
  \begin{bmatrix}
    \hat{G} & \hat{F} \\
    - \undertilde{\hat{F}} & - \undertilde{\hat{G}}
  \end{bmatrix}_{(\bm{k}, \epsilon)} , \ 
  \check{G}^A (\bm{k}, \epsilon) 
  = 
  \left[
    \check{G}^R (\bm{k}, \epsilon)
  \right]^{\dagger} .\label{eq:greenfunction}
\end{align} 
The thermoelectric coefficient in a spin-singlet $s$-wave superconductor
is calculated to be
\begin{align}
  \alpha 
  &=
  \alpha_{\mathrm{N}} 
  \frac{3}{2 \pi^2}
  \int_{\Delta/T}^{\infty} dx 
  \frac{x^2}{\cosh^2 (x/2)} , \\
  \alpha_{\mathrm{N}} 
  &=
  \frac{\pi^2}{3 d} e \hbar T \tau C_0 , \ \ \ 
  C_0 
  = \frac{d}{d \xi} N(\xi) v^2(\xi) \Big|_{\xi = 0} ,
\end{align}
where $\tau$ is the relaxation time due to random impurities.
The results are identical to the 
previous results~\cite{galperin1974thermoelectric}.
The lower limit of integral reflects the effects of superconductivity.
The thermoelectric coefficient of an $s$-wave superconductor is 
exponentially smaller than $\alpha_{\mathrm{N}}$ at low temperatures 
$\alpha \propto \alpha_{\mathrm{N}} \exp(- \Delta/T) \ll \alpha_{\mathrm{N}}$.

\end{widetext}

\begin{figure*}[tttt]
  \centering
  \includegraphics[width = 2\columnwidth]{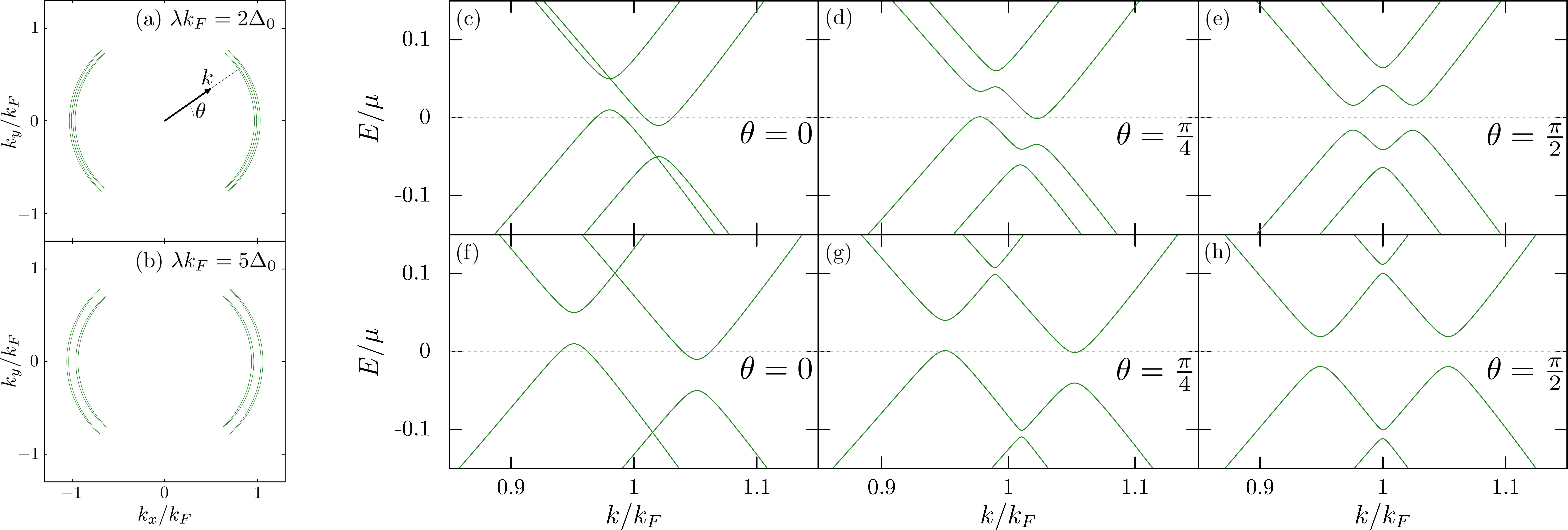}
  \caption{
    The BFSs at $V = 1.5 \Delta_0$ are displayed 
	for $\lambda k_F = 2\Delta_0$ in (a) and for  $\lambda k_F = 5\Delta_0$ in (b). 
	The dispersion along several directions in momentum space are plotted 
	for  $\lambda k_F = 2\Delta_0$ in (c)-(e) and for  $\lambda k_F = 5\Delta_0$
 in (f)-(h). 
    The vertical axis is normalized to the chemical potential $\mu$.
  }
  \label{fig:energyspe}
\end{figure*}
\section{Energy dispersion}\label{AppB}

In this Appendix, we discuss the role of SOI in the formation of BFSs.
Figs.~\ref{fig:energyspe} (a) and (b) show the BFSs in momentum space 
at $V = 1.5 \Delta_0$ calculated for 
$\lambda k_F = 2 \Delta_0$ and $\lambda k_F = 5 \Delta_0$, respectively.
When $\lambda k_F$ increases, 
the outer BFSs and the inner BFSs shift the opposite direction to each other 
along the $k_x$ axis.
The size of the BFSs in the $k_y$ direction is insensitive to $\lambda k_F$. 
A direction in the two-dimensional momentum space is specified by an angle $\theta$.
The energy dispersions are plotted as a function of the wavenumber for several 
directions in momentum space.  
The results for $\lambda k_F = 2 \Delta_0$ are displayed in Figs.~\ref{fig:energyspe} (c)-(e) and
those for $\lambda k_F = 5 \Delta_0$ are displayed in Figs.~\ref{fig:energyspe} (f)-(h).
Comparing (c)-(e) with (f)-(h), 
the SOI shifts the energy dispersions in the $k$-direction, which is the main role 
of the SOI in the formation of BFSs. 
As a result, 
the characteristic features of the thermoelectric effect 
are insensitive to $\lambda k_F$.

\section{Thermoelectric effect in a \texorpdfstring{$d$}{d}-wave superconductor with BFSs}\label{AppC}
\begin{figure}[t]
  \includegraphics[width=8cm]{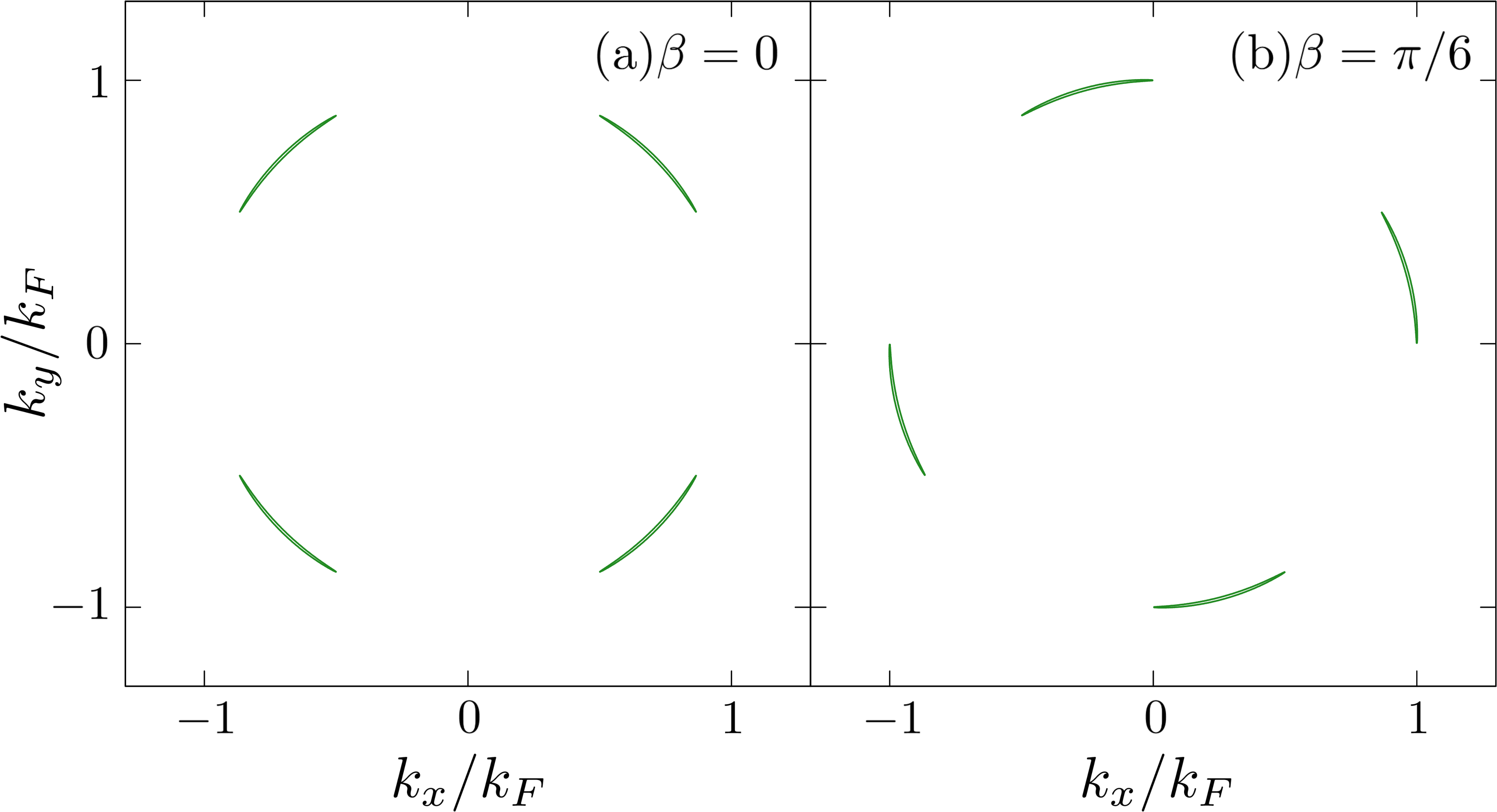}
  \caption{
    The BFSs in a $d$-wave superconductor at $V = 0.5 \Delta_0$ 
	  are displayed  for $\beta = 0$ in (a) and for $\beta = \pi/6$ in (b), respectively.
  }
  \label{fig:BFS_dwave}
\end{figure}

In this Appendix, we discuss the thermoelectric effect in a $d$-wave superconductor with the BFSs,
and compare our results with the results in a normal-metal/superconductor junction~\cite{Pal:prb2024}.
The BFSs also exist by the inflation of point nodes in a $d$-wave superconductor in two-dimension~\cite{Yang:prb1998}.
The BdG Hamiltonian in a $d$-wave superconductor under a Zeeman field is represented as
\begin{align}
  \check{H}_{\mathrm{BdG}}
  &=
  \begin{bmatrix}
    \hat{h}_{\mathrm{N}} (\bm{k}) & \Delta_\theta i \hat{\sigma}_y \\
    - \Delta_\theta \hat{\sigma}_y & - \hat{h}^\ast_{\mathrm{N}} (-\bm{k})
  \end{bmatrix} , \\
  \hat{h}_{\mathrm{N}} &= \xi_{\bm{k}} \hat{\sigma}_0 - \bm{V} \cdot \hat{\bm{\sigma}} , \\
  \Delta_\theta
  &=
  \Delta \cos\big[ 2(\theta + \beta) \big] ,
\end{align}
with $\theta = \tan^{-1} (k_y / k_x)$ and orientation angle $\beta$, 
where a Zeeman field is applied in the $y$ direcion.
The energy eigenvalues are calculated to be
\begin{align}
  E_{\bm{k}} = \sqrt{\xi_{\bm{k}}^2 + \Delta_{\theta}^2} \pm V , \quad - \sqrt{\xi_{\bm{k}}^2 + \Delta_{\theta}^2} \pm V.
\end{align}
Fig.~\ref{fig:BFS_dwave} shows the BFSs (possible eigenvalues of the BdG Hamiltonian at zero energy) 
for $\beta = 0$ ($d_{x^2 - y^2}$-wave) in (a) and those for $\beta = \pi/6$ in (b), 
where we fix a Zeeman potential at $V = 0.5 \Delta_0$.
The BFSs appear around the point nodes even for $V < \Delta_0$.

\begin{figure}[t]
  \includegraphics[width=7cm]{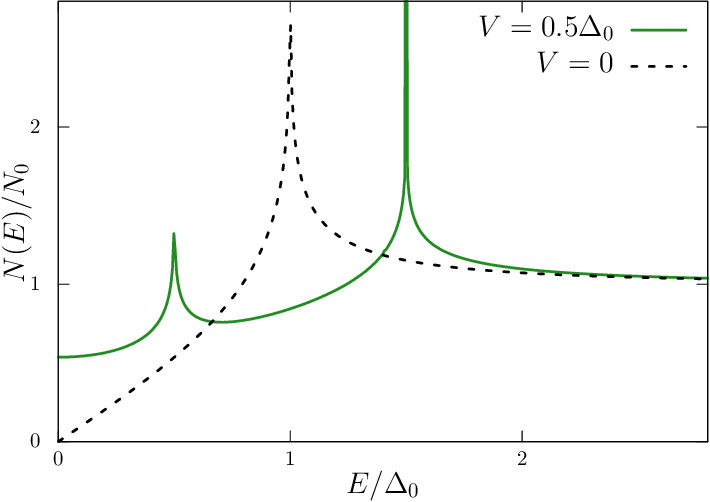}
  \caption{
    The density of states (DOS) in a $d$-wave superconductor.
    The solid line shows the DOS at $V = 0.5 \Delta_0$ and the dotted line shows the DOS at $V = 0$.
  }
  \label{fig:dos_dwave}
  \includegraphics[width=8.5cm]{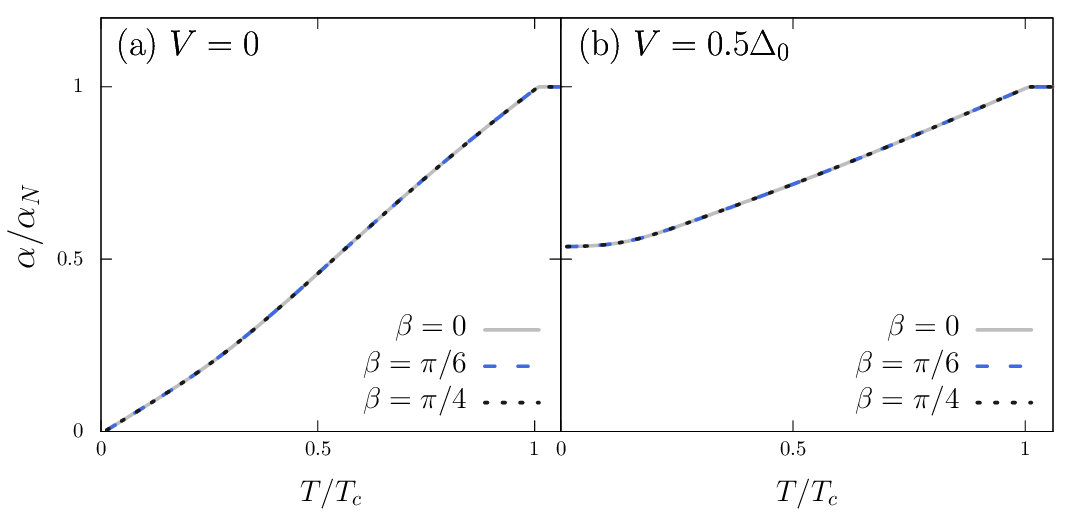}
  \caption{
    The thermoelectric coefficient versus temperature in a $d$-wave superconductor 
    for $V=0$ and $V = 0.5\Delta_0$ are displayed in (a) and (b), respectively.
    The coefficients for $\beta = 0$, $\beta = \pi/6$, and $\beta = \pi/4$ 
    are represented with the solid line, dashed line, and dotted line.
    } 
  \label{fig:thermoele_dwave}
\end{figure}

The DOS in a $d$-wave superconductor with the BFSs calculated to be
\begin{align}
  N(E)
  &=
  \frac{N_0}{2} \int_{0}^{2 \pi} \frac{d \theta}{2 \pi}
  \Bigg[
    \frac{|E + V| \, \Theta (|E + V| - |\Delta_{\theta}|)}{\sqrt{(E + V)^2 - \Delta_{\theta}^2}}  \nonumber \\
    &\qquad +
    \frac{|E - V| \, \Theta (|E - V| - |\Delta_{\theta}|)}{\sqrt{(E - V)^2 - \Delta_{\theta}^2}}  
  \Bigg] , 
\end{align}
where $\Theta$ is the step function.
In Fig.~\ref{fig:dos_dwave}, we present the DOS in the presence of the BFSs in a $d$-wave superconductor.
The DOS vanishes at $E=0$ in the absence of a Zeeman field, whereas 
that remains a finite value for $V=0.5\Delta_0$.

The shape of the pair potential on the Fermi surface is an important factor 
in a thermoelectric effect.
Because $|\Delta_{\theta}| = |\Delta_{\theta + \frac{\pi}{2}}|$ holds in a $d$-wave superconductor,
the thermoelectric coefficients in the $x$ direction and that in the $y$ direction are 
equal to each other.
Therefore, we calculate the thermoelectric coefficients in the $x$ direction 
$\alpha^x$ for several $\beta$.
The thermoelectric coefficient is represented as
\begin{align}
  \alpha^x 
  &=
  \frac{e \hbar}{32 \pi T^2}
  \int_{-\infty}^{\infty} d \epsilon
  \frac{\epsilon}{\cosh^2 (\epsilon/2T)} \nonumber \\
  &\quad \times
  \int_{0}^{2 \pi} \frac{d \theta}{2 \pi} \cos^2 \theta 
  \int_{-\infty}^{\infty} d \xi N(\xi) v^2 (\xi) I(\xi, \theta, \epsilon) .
  \label{eq:alphax}
\end{align}
The results of the thermoelectric coefficients are plotted 
as a function of temperature in Fig.~\ref{fig:thermoele_dwave}.
The results for $V = 0$ and those for $V = 0.5\Delta_0$ are shown in (a) and (b), respectively.
We assume that the dependence of $\Delta$ on temperatures is described by BCS theory 
and is common in (a) and (b).
The solid, dashed and dotted lines represent the coefficients for $\beta = 0$, $\beta = \pi/6$ and $\beta = \pi/4$ respectively.
The numerical results indicate that the thermoelectric coefficients 
in a $d$-wave superconductor are isotropic in real space.
To make this point clear, we analyze the integral of $\theta$ in Eq.~\eqref{eq:alphax}.
As $\theta$ dependence of $I(\xi,\theta,\epsilon)$ is derived from 
$|\Delta_{\theta+\beta}|$ in Eq.~\eqref{eq:gba1}, the relation
\begin{align}
  I(\xi,\theta,\epsilon) = I(\xi,\theta+\pi/2,\epsilon)
\end{align} 
holds in a $d$-wave superconductor even in a Zeeman field.
The integral of $\theta$ is calculated to be
\begin{align}
  &\int_{0}^{2\pi} d \theta \, \cos^2 \theta \, I(\xi,\theta+\beta,\epsilon) \nonumber \\
  &\quad = 2 \int_{0}^{\pi} d \theta \, \cos^2 (\theta - \beta) \, I(\xi,\theta,\epsilon), \nonumber \\
  &\quad = 2 \int_{0}^{\frac{\pi}{2}} d \theta \, \cos^2 (\theta - \beta) \, I(\xi,\theta,\epsilon) \nonumber \\
  &\qquad + 2 \int_{\frac{\pi}{2}}^{\pi} d \theta \, \cos^2 (\theta - \beta) \, I(\xi,\theta,\epsilon), \nonumber \\
  &\quad = 2 \int_{0}^{\frac{\pi}{2}} d \theta \, \cos^2 (\theta - \beta) \, I(\xi,\theta,\epsilon) \nonumber \\
  &\qquad + 2 \int_{0}^{\frac{\pi}{2}} d \theta \, \cos^2 (\theta - \beta + \pi/2) \, I(\xi,\theta+\pi/2,\epsilon), \nonumber \\
  &\quad = 2 \int_{0}^{\frac{\pi}{2}} d \theta \,
  \left\{ \cos^2 (\theta - \beta) + \sin^2 (\theta - \beta) \right\} \, I(\xi,\theta,\epsilon), \nonumber \\
  &\quad = 2 \int_{0}^{\frac{\pi}{2}} d \theta \, I(\xi,\theta,\epsilon) \quad (\text{independent of $\beta$}).
\end{align}
Therefore, the thermoelectric coefficients are independent of $\beta$ in a $d$-wave superconductor.
The results also indicate $\alpha^x < \alpha_N$.
To realize a large thermoelectric effect, a large Zeeman field $V \approx \Delta$ is necessary to 
shift the coherence peak in DOS to zero energy.
However, a superconducting state becomes unstable in such large Zeeman fields.

In a normal-metal/$d$-wave superconductor junction, the thermoelectric effect shows the anisotropy 
and the thermoelectric coefficient in a certain direction can be larger than that in the normal state.
Such behaviors originate from the Andreev bound states at the interface of the 
junction~\cite{Pal:prb2024}.

\end{document}